\documentclass[pra,twocolumn]{revtex4}
\usepackage{epsfig, amsfonts, mathrsfs, amsbsy}

\newcommand{\ket}[1]{\mbox{\ensuremath{|#1\rangle}}}
\newcommand{\bra}[1]{\mbox{\ensuremath{\langle#1|}}}
\newcommand{\bm}[1]{\mbox{\bf{#1}}}

\graphicspath{{figs/}}

\begin{document}

\title{Interaction of light with a single atom in the strong focusing
regime}
\author{Syed Abdullah Aljunid$^1$, Brenda Chng$^1$, Martin Paesold$^{1,2}$, 
  Gleb Maslennikov$^1$ and Christian Kurtsiefer$^1$}
\email[]{christian.kurtsiefer@gmail.com}
\homepage[]{http://www.quantumlah.org}

\affiliation{$^1$Centre for Quantum Technologies / Department of Physics,
National University of Singapore, Singapore}
\affiliation{$^2$ETH Z\"urich, Switzerland}



\date{\today}

\begin{abstract}
We consider the near-resonant interaction between a single atom and a focused
light mode, where a single atom localized at the focus of a lens can scatter a
significant fraction of light. Complementary to previous experiments on
extinction and phase shift effects of a single atom, we report here on the
measurement of coherently backscattered light. The strength of the
observed effect suggests combining strong focusing with the
well-established methods of cavity QED. We consider theoretically a
nearly concentric cavity, which should allow for a strongly focused optical
mode. Simple estimates show that in a such case one can expect a significant
single photon Rabi frequency. This opens new perspectives and a
possibility to scale up the system consisting of many atom+cavity nodes for
quantum networking due to a significant technical simplification of the
atom--light interfaces.
\end{abstract}

\maketitle

\section{Introduction}
In order to manage more complex tasks in quantum communication and information
processing, it became obvious in the recent years that different physical
systems have different strengths in various quantum information processing
primitives. Thus, different quantum systems have to be combined for more
complex tasks, and interfaces between various microscopic systems and photons
in the optical domain --- as quantum information carriers which allow easy
decoherence-free transport of quantum information --- are of strong interest. 

To accomplish a substantial information exchange between photons and
microscopic systems like atoms, or to implement an atom-mediated
interaction between flying qubits, a strong interaction between the photons
and atoms is essential. So far, the canonical approach to this problem
in the makes use of cavities with a high finesse and a small mode
volume. This significantly enhances the electrical field of a single photon such
that the atomic state could be substantially affected \cite{Kimble:98}.
This technique has been proven to be amazingly
successful \cite{Ye:99,Mabuchi:02,Armani:03,Hennrich:05}, but it remains a
technological challenge to work with such cavities. Despite the
dramatic success in increasing resonator finesse with smart
dielectric structures \cite{Armani:03}, the bulk of
cavity QED experiments in the optical domain still makes use of cavities based
on sophisticated dielectric coatings \cite{Hood:01,Akoi:06,Steinmetz:06},
which are still an art to make. This, together with the challenge to stabilize
the cavities against vibrations make it seem hard to scale up such atom--light
interfaces to the desired quantum networks \cite{Cirac:97}.

\begin{figure} 
  \centerline{\epsfxsize65mm \epsffile{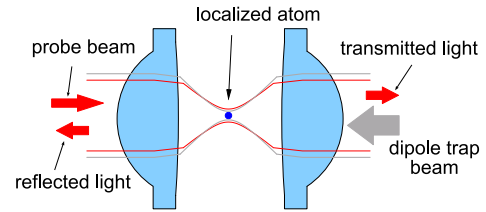}}
  \caption{\label{fig:concept}Concept of an atom interacting with a Gaussian
    light mode using a pair of lenses.}
\end{figure}

An alternative approach for increasing the electromagnetic field at the
location of an atom 
considers strong focusing of the optical mode.
This technique has been applied to observe strong
interaction of the electromagnetic fields with molecules
\cite{molecule1,molecule2}, 
quantum dots \cite{Vamivakas:07}, and single atoms \cite{Tey:08,Syed:09}
as possibly one of the cleanest physical systems for quantum information
processing. The basic configuration of such an interaction scheme is shown in
figure~\ref{fig:concept}. Efforts are under way to reach near perfect coupling
between field modes and electrical dipole transitions in ions
\cite{Sondermann:07}.

In this paper, we report on measurements of the
reflection of light in a tightly focused mode by a single atom in section
\ref{sec:experiment}, complementing the observation of transmission
\cite{Tey:08} and phase shift \cite{Syed:09}.
The essence of such experiments can be understood by treating the light field
classically, since the focusing geometry only modifies the boundary
conditions of the electromagnetic field.

In order to investigate the interaction between light and the atom on the
single photon level, however, we need to move on to a quantized description of
the electromagnetic field. In section \ref{sec:quantization}, we describe how
the effects of the boundary conditions in a strong focusing regime can be integrated
into a quantized description of the electromagnetic field. We consider both
running wave geometries, which connect to `flying qubits', i.e., for a
continuous mode spectrum of the field, and a discrete mode spectrum as it
may be found in resonator configurations. The strong enhancement of the light
field due to focusing may open a complementary route to bring 
atom--light interaction into the strong coupling regime, which for now was
achievable only with the aid of small cavities with a high finesse. In section
\ref{sec:cavity}, we follow up on earlier ideas of cavities with a small focus
\cite{Morrin:94,Haase:06} and estimate the coupling strength in a
Jaynes--Cummings model of atom--field interaction for a cavity which combines
our strong focusing results with a resonator.

\section{Experimental observation of backscattering from a single atom}\label{sec:experiment}
To complement our earlier measurements on the transmission of a light beam
focused on an atom and the atom-induced phase shift, we measured the
reflection of light into the source mode by the atom. Such a
reflectivity has been shown to be perfect in the idealized case of an
electrical field that exactly matches a directional dipole wave
\cite{zumofen:2008}. However, such an electric field is not easily
prepared in practice, so light fields with a Gaussian envelope are commonly
used instead, 
as they emerge in a good approximation from optical fibers. There, the
maximal reflectivity for circularly polarized light is limited to about 53\% \cite{Tey:09}.

\begin{figure} \centerline{\epsfxsize86mm 
\epsffile{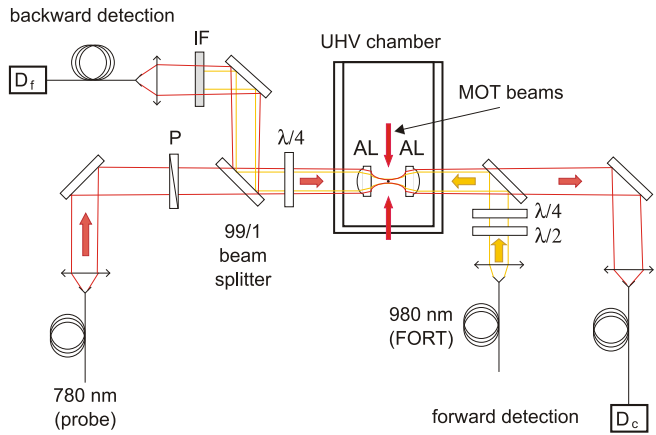}} 
  \caption{\label{fig:setup} Experimental setup for measuring
    extinction/reflection. A single atom is localized in the focus of two
    lenses with a far-off resonant trap, and exposed to the focused probe
    light field. Transmission and reflection measurements are carried out with
    single photon detectors behind single mode fibers in forward and backward
    direction.}  
\end{figure}

A schematic outline of our experiment is shown in figure~\ref{fig:setup}. A
single $^{87}$Rb atom is localized at the focus of two aspheric lenses with a
circularly polarized far-off
resonant dipole trap (FORT) with an operating wavelength of 980\,nm in an
ultrahigh vacuum chamber. Prior to localizing the atom, its kinetic energy is
lowered to 15--30\,$\mu$K 
with the aid of a magneto-optical trap and a
subsequent molasses cooling step. A probe beam emerging from a single mode
optical fiber with a wavelength of $\lambda=$780\,nm is prepared into a
circular polarization state, and sent through the same
lens pair in a counter-propagating direction. Its frequency can be tuned over
the transition from the $5S_{1/2}, F=2$ electronic ground state to the $5P_{3/2},
F=3$ excited state, 
and has a waist parameter of $w_L$=1.25\,mm just before the focusing lens with
$f=4.5$\,mm. The transmitted light is collimated by an identical
lens, and coupled into another single mode optical fiber, from where it
reaches a photon counting module, D$_c$. To capture the reflected light, a 99:1
beam splitter is inserted into the probe beam prior to the lens
arrangement, such that most of 
the reflected light can be directed into another single mode optical fiber and
detected by a second single photon detector, D$_f$. Dichroic mirrors and
an interference filter are used to suppress the residual trap light
propagating in this direction. Detector D$_f$ is also used to observe the
fluorescence from an atom in the FORT during the loading process from the
magneto-optical trap. To avoid optical pumping into the $F=1$ ground state,
re-pump light at $\lambda=795$\,nm resonant with the D1 line is used to
transfer the atom back to the approximate two-level system via the $5P_{1/2},
F=2$ level. Details of the loading and state preparation sequence
can be found in \cite{Tey:08}.

\begin{figure} \centerline{\epsfxsize86mm 
\epsffile{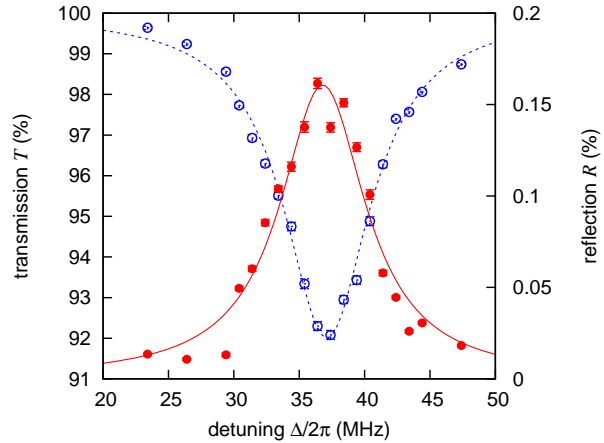}}
  \caption{\label{fig:expresult} Combined results for transmission and
    reflection from the atom--lens setup as a function of the detuning with
    respect to the unperturbed atomic transition. For this focusing strength,
    we observe a maximal reflection of 0.17\% of the incoming power (open
    circles), coinciding with the minimal transmission of 92\% (filled
    circles).}
\end{figure}

The results from the transmission and reflection measurement obtained by
sweeping the frequency of the probe beam across the resonance are shown in
figure~\ref{fig:expresult}. The transmission results are
obtained by normalizing the events observed in detector D$_c$ for a particular
detuning in the presence of an atom to events in a time interval without the
atom. Similarly, the reflectivity is obtained by normalizing
the events detected with D$_f$ against the normalization events in detector D$_c$
without an atom in the trap. On resonance, the number of events observed in
detector D$_f$ was about 13\,s$^{-1}$ above the detector background
(100--400\,s$^{-1}$), and the data points shown were obtained
by collecting events over an integration time of 50 minutes per point. 
 Error bars depicted
in the graph represent propagated Poissonian counting statistics into
the final reflection/transmission value.

Both transmission and reflection measurements seem to follow the expected
Lorentzian line profile. A best fit of our data to such a profile (solid lines
in figure~\ref{fig:expresult}) with center detuning
$\Delta f_0$, full with of half maximum (FWHM) $\delta f$, and amplitude
as free parameters for the transmission measurement results in $\Delta
f_0=37.1\pm0.1$\,MHz and $\delta f=8.1\pm0.3$\,MHz, and a maximal extinction
value of $1-T=8.2\pm0.2$\%. For the reflection measurement, we find
$\Delta f_0=36.8\pm0.2$\,MHz, $\delta f=8.0\pm0.6$\,MHz, and a peak reflectivity
of $0.161\pm0.007$\%. 

The reflection/transmission measurements show complementary
dependency on the detuning within our experimental uncertainty. The
center frequency of the resonance line is shifted to the blue due to the AC
Stark shift induced by the optical dipole trap, and the observed
widths are slightly larger than the natural line width of the probed D2 line
(6.06\,MHz). 

As shown in \cite{Tey:09}, the transmission and reflection properties
can be derived from a dimensionless quantity 
\begin{equation}
R_{sc}(u)={3\over4u^3}e^{2/u^2}\left[\Gamma(-{1\over4},{1\over u^2})+
  u\Gamma({1\over4},{1\over u^2})\right]^2\,,
\end{equation}
which for a small focusing parameter $u=w_L/f$ has the interpretation of a
scattering probability of a photon from the atom, and exhibits a maximum of
about 1.4 for $u\approx2$. 
The focusing parameter itself is connected to the Gaussian beam divergence
$\Theta$ via $\tan\Theta=u$.
For collection into modes overlapping with the excitation mode, the
extinction $1-T$ and reflectivity $R$ are given by
\begin{equation}
1-T=1-\left|1-{R_{sc}\over2}\right|^2\quad\textrm{and}\quad R=R_{sc}^2/4\,.
\end{equation}
The amplitude coefficient of the fit to the transmission measurement for this
model corresponds to $R_{sc}=0.083\pm0.002$, and the reflectivity result to
$R_{sc}=0.080\pm0.002$. Both values are in agreement within experimental
uncertainties. In our experiment, however, we used a focusing parameter
$u=0.278$ corresponding to $R_{sc}=0.201$. Apart from lens imperfections, a
very likely 
explanation for this discrepancy is the delocalization of the atom from the
focus due to its residual kinetic energy, which would reduce 
both the electrical field felt by the atom as well as the overlap of the
radiated field with the receiving modes. A quantitative numerical estimation
of this effect is in preparation \cite{colin:10}.

From this experiment, it seems that
the transmission and reflection properties for light in a
strongly focused mode are well understood, and motivates further investigation
of atom--light interaction in the strong focusing regime.

\section{Field quantization for focused Gaussian modes}\label{sec:quantization}
The field quantization of Gaussian beams is relatively straightforward and has
been done many times in the paraxial regime \cite{meystre:99}. We quickly
revisit one approach here, mostly to clarify the nomenclature we use later on
for the strongly focused scenario. For simplicity, we begin with periodic
boundary conditions and a `quantization length' $L$.

We start with an expression for the electric field operator in a generalized
mode decomposition,
\begin{equation}\label{eq:generalfieldoperator}
  \hat{\bm{E}}(\bm{x},t)=i\sum\limits_j\mathscr{E}_{\omega_j}
  \left[\bm{g}_j(\bm{x})\hat{a}_j(t) -
    \bm{g}_j^*(\bm{x})\hat{a}_j^\dagger(t)
  \right]\,,
\end{equation}
where the spatial dependency and polarization property of a mode $j$ is
covered by a mode function $\bm{g}_j(\bm{x})$ and the time dependency by the
ladder operators $\hat{a}_j, \hat{a}_j^\dagger$. The dimensional components,
together with a normalization of the mode function, 
are lumped into the constant $\mathscr{E}_{\omega_j}$.

The free field Hamiltonian of the total field
\begin{equation}
\hat{H}_0={\epsilon_0\over2}\int\!d\bm{x}\,\left[\hat{\bm{E}}^2(\bm{x}) + c^2 \hat{\bm{B}}^2(\bm{x})\right]
\end{equation}
then can be written as a sum of contributions from different modes $j$:
\begin{equation}
\hat{H}_0=\sum\limits_j\hat{H}_j=
\sum\limits_j{\epsilon_0\over2}\int\!d\bm{x}\,\left[\hat{\bm{E}_j}^2(\bm{x}) + c^2 \hat{\bm{B}_j}^2(\bm{x})\right]\,.
\end{equation}
For a given mode $j$, the mode function $\bm{g}_j(\bm{x})$ has to be
compatible with the Maxwell equations for an angular frequency of $\omega_j$,
and the normalization constant $\mathscr{E}_{\omega_j}$ needs to be chosen
that $\hat{H}_j$ coincides with the Hamiltonian of a harmonic oscillator. This
leads to 
\begin{equation}
\mathscr{E}_{\omega_j}=\sqrt{\hbar\omega_j\over2\epsilon_0 V_j}\,,
\end{equation}
with an effective mode volume $V_j$ given by
\begin{equation}\label{eq:modevolume}
V_j:={1\over2}\int\!\!d\bm{x}\,\left[\left|\bm{g}_j(\bm{x})\right|^2 +
  {c^2\over\omega_j^2}\left|\nabla\times\bm{g}_j(\bm{x})\right|^2\right].
\end{equation}
In vacuum, electric and magnetic contributions to the total field energy are
the same, such that we can reduce the above integral to the simplified
expression 
\begin{equation}\label{eq:modevolume2}
V_j=\int\!\!d\bm{x}\,\left|\bm{g}_j(\bm{x})\right|^2\,.
\end{equation}

In the following, we now restrict the treatment of the electromagnetic field
to a single mode $j$ of the electromagnetic field, and drop the mode index.
\begin{figure}
\centerline{\epsfxsize86mm\epsffile{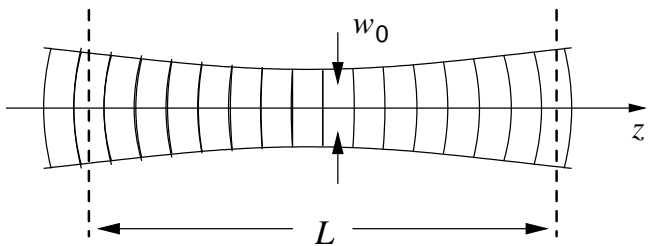}}
\caption{\label{fig:gaussquanti}
  Quantization geometry of a Gaussian beam of a waist $w_0$. To
  discretize the mode longitudinal mode index, we introduce periodic boundary
  conditions with a quantization length $L$.}
\end{figure}
In paraxial optics, a Gaussian beam is characterized by the waist $w_0$ and
the longitudinal mode index, usually described by a wave number $k_j$. In a
regime where there is no significant wavefront curvature (i.e., $L$ is smaller
than the Rayleigh range $z_R=\pi w_0^2/\lambda$), the mode function is given by
\begin{equation}\label{eq:gaussmodefunction}
  \bm{g}(\bm{x})=\pmb{\varepsilon}e^{-\rho^2/w_0^2}e^{ikz}\,,
\end{equation}
with a radial distance $\rho$, a transverse polarization vector
$\pmb{\varepsilon}$, and a position $z$ along the
propagation direction. The dispersion relation for this mode is given by
$\omega^2=c_0^2(k^2+2/w^2)$.
The spatial integration in eq.~(\ref{eq:modevolume2}) is well-defined if the
volume in propagation direction $z$ is restricted to the quantization length
$L$ (see Fig.~\ref{fig:gaussquanti}), leading to a result for the effective
mode volume of
\begin{equation}
V=\pi L w^2/2\,,
\end{equation}
and consequently to a normalization constant $\mathscr{E}$ of
\begin{equation}\label{eq:gaussnormfield}
  \mathscr{E}=\sqrt{\hbar\omega\over \pi w^2 L \epsilon_0}\,.
\end{equation}
This expression is valid in  a regime where the variation of the beam waist
$w$ along the interval in $z$ direction does not change, i.e., the
quantization length $L\ll z_R$. It turns out that this restriction is actually
not necessary, and we can easily extend this result to focused beams.

\subsection{Strong focusing limit}
To consider focused Gaussian beams, we  keep in mind that even in a strong
focusing regime ($u$ large), light propagates either in $+z$ or $-z$
direction, i.e., all the optical power transported in the mode has to pass
through each transverse plane. Hence,  the contribution to the volume integral
for each slice $dz$ in eq.~(\ref{eq:modevolume}) is
independent of $z$. Attention only has to be payed {\em where} the modulus of
$\bm{g}(\bm{x})$ is referenced to 1. For the case $L\ll z_R$ treated above, the
mode function can be chosen to $|\bm{g}(\bm{x})|=1$ on the entire optical
axes. If one includes any focusing of the Gaussian mode, $|\bm{g}(\bm{x})|$
varies along the optical axes, and will show a maximum in the focus. In a
regime where the paraxial approximation is still valid, 
we can simply use the waist $w=w_f$ in the focus in
eq.~(\ref{eq:gaussnormfield}) to obtain the correct normalization constant.

In the strong focusing regime where the paraxial approximation
breaks down, a slightly different approach has to be used to estimate the
electrical field at the location of the atom. Again, we start with a
Gaussian 
mode at a location far away from the focus; for convenience, we consider a
configuration as depicted in figure~\ref{fig:concept}. We assign a Gaussian
beam waist $w_L$ at the collimated part of the beam outside the lens pair, and
choose the mode function such that there, $|\bm{g}(\bm{x})|=1$ on the optical
axis. The ratio of the electrical field amplitude on the axis before the lens
pair, $E_L$, and in the focus of the system, $E_A$, can be
expressed in a closed form for circularly polarized light. This
ratio is determined by the numerical aperture of the focusing
setup, which for Gaussian beams is better described by the focusing parameter
$u=w_L/f$. One can show \cite{Tey:09} that
\begin{equation}\label{eq:fieldenhancement}
\left({E_A\over E_L}\right)^2={\pi^2 w_L^2R_{sc}(u)\over3\lambda^2}\,.
\end{equation}

We now can take the normalization expression from
eq.~(\ref{eq:gaussnormfield}) with a waist $w_L$ before the lens, and multiply
it with the field ratio determined by eq.~(\ref{eq:fieldenhancement}) due to
the focusing; we arrive at
\begin{equation}\label{eq:focusnormalize}
\mathcal{E}=\sqrt{\pi\hbar\omega R_{sc}(u)\over 3 \lambda^2 L \epsilon_0}\,.
\end{equation}
The normalization constant now only depends on the wavelength, quantization
length and focusing strength (or divergence) of the mode.

\subsection{Wave packets}
To include wave packets of light in a quantized field
description, a continuous mode decomposition
is helpful, in which `localized photons' can be expressed as a superposition
of contributions from a continuous one-dimensional spectrum of modes, labeled
by a continuous longitudinal wave number $k=\omega/c$, or the angular
frequency $\omega$ directly. Our treatment above is valid for such a case if
we apply the simple transition $L\rightarrow\infty$, and replace the
discrete sum over modes e.g. in eq.~(\ref{eq:generalfieldoperator}) by an
integral: 
\begin{equation}
\sum_j\rightarrow {L\over2\pi c}\int\!d\omega
\end{equation}
The expression for the normalization constants $\mathcal{E}$ still can be
taken from the expressions above, and in all physically meaningful
quantities the length $L$ should not appear anymore.
Such a field decomposition allows the treatment of wave packets of various
shapes. In particular, it allows to understand the dynamics of the
absorption of single photon states \cite{Sondermann:07,Stobinska:09} or
pulsed coherent states of the field.

\subsection{Combination with optical cavities}
The introduction of a quantization length $L$ for the field quantization
has the big advantage that the mode spectrum of the electromagnetic field
becomes discrete, and interaction of an atom with a single field mode can be
investigated. Among other reasons, a discrete mode spectrum allows for a
reasonably clean definition of what is meant by a single photon; for a
continuum of modes, this is more ambiguous.

The field-enhancing effect of a small focus (`hourglass modes') has been
identified as a method for enhancement of the electrical field in an
optical cavity at the location of atoms before \cite{Morrin:94,Haase:06}.
Here, we would like to extend the description of the electrical field to
strongly focused modes.

The simplest transition from the treatment presented above to a resonant
structure involves a ring cavity, which reflects exactly the periodic boundary
conditions above. Thus, we can directly use eq.~(\ref{eq:focusnormalize}) for
the normalization constant $\mathcal{E}$ for this case.

For the more conventional cavities with a standing wave field mode,
quantization can be adapted from the case of periodic boundary conditions by
modifying the propagating part of the mode: the $e^{ikz}$ term has to be
replaced with the standing wave component $\sin(kz)$ for a proper choice of
origin, and $k=N\pi/L$ with an integer  $N$. Compared to periodic boundary
conditions, the $\sin^2(kz)$ modulation in the field energy density reduces the
effective mode volume by a factor of 2, 
and the normalization constant $\mathcal{E}$ is given by 
\begin{equation}
  \mathcal{E}=\sqrt{2\pi\hbar\omega R_{sc}(u)\over 3 \lambda^2 L \epsilon_0}\,.
\end{equation}

For a standing wave resonator, the field amplitude has also a strong variation
in longitudinal direction. To ensure a strong coupling of an atom with the
resonator mode, the atom needs to be localized close to the anti-node of the
standing wave.

\section{Estimation of coupling parameters}\label{sec:cavity}
\begin{figure}
\centerline{\epsfxsize86mm\epsffile{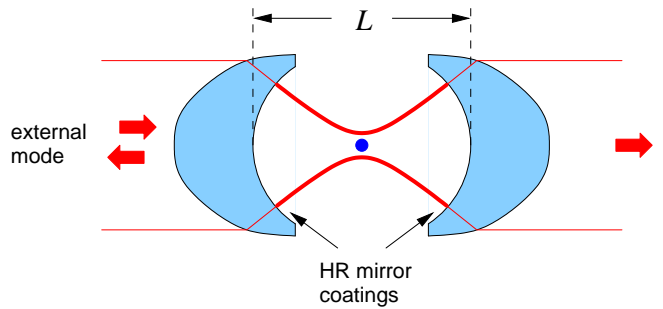}}
\caption{\label{fig:cavityconcept}
  Concept of an atom in a cavity with a strongly focused resonator mode and a
  cavity length $L$, coupling to collimated running modes with a pair of
  anaclastic lenses.}
\end{figure}

Since a large fraction of the scientific work on strong atom--light coupling is
carried out where field modes are supported by a cavity with a discrete mode
spectrum, we estimate quantitatively the combination of the field
enhancement in a cavity by strong focusing, as shown
in figure~\ref{fig:cavityconcept}. Similarly to \cite{Morrin:94}, two
spherical surfaces with a separation $L$ form a cavity such that the cavity
mode corresponds to a the focused light field configuration in the above
discussed strong focusing regime, and is characterized by a focusing parameter
$u$. We assume that
the cavity mode has still a Gaussian transverse envelope, although for a
strong focusing regime, the cavity is close to the edge of the stability region.

\subsection{Interaction Hamiltonian}
With the electrical field operator for an optical
mode overlapping with an atom we can easily obtain the coupling strength
necessary for a treatment of a two-level atom and the resonant field in a
Jaynes--Cummings model. The interaction Hamiltonian of the field with an atomic
electric dipole moment $\bm{d}$ is given by 
\begin{eqnarray}
\hat{H}_I&=&\hat{\bm{E}}_A\cdot\hat{\bm{d}} \nonumber \\
&=&i\mathcal{E}d_{\textrm{eff}}\left(\ket{e}\bra{g}\hat{a}g(\bm{x}_A)-\ket{g}\bra{e}\hat{a}^{\dagger}g^*(\bm{x}_A)\right)
\nonumber \\
&=:&ig_0\left(\ket{e}\bra{g}\hat{a}-\ket{g}\bra{e}\hat{a}^{\dagger})\right)\,,
\label{eq:jaynescummings}
\end{eqnarray}
where we choose $g(\bm{x}_A)=1$ at the location of the atom. We used an
effective dipole matrix element $d_{\textrm{eff}}$, which, for a field
polarization matching the transition between the two levels and a unit
Clebsch-Gordan coefficient, 
is related to the decay time constant $\tau$ for the atomic transition by
\begin{equation}
d_{\textrm{eff}}=\sqrt{3\epsilon_0h\lambda^3\over8\pi^2\tau}\,.
\end{equation}
Therein, $h$ is the Planck constant and $\lambda$ the vacuum wavelength
corresponding to the optical transition between ground- and excited state
\cite{CCT:98}.
Usually, the coupling strength $g_0=\mathcal{E}d_{\textrm{eff}}$ is quoted to
characterize an atom+cavity system, which for a standing wave cavity is given
by 
\begin{equation}
  g_0=\hbar\sqrt{\pi c R_{sc}(u)\over\tau L}\,.
\end{equation}

\begin{figure}
\centerline{\epsfxsize86mm\epsffile{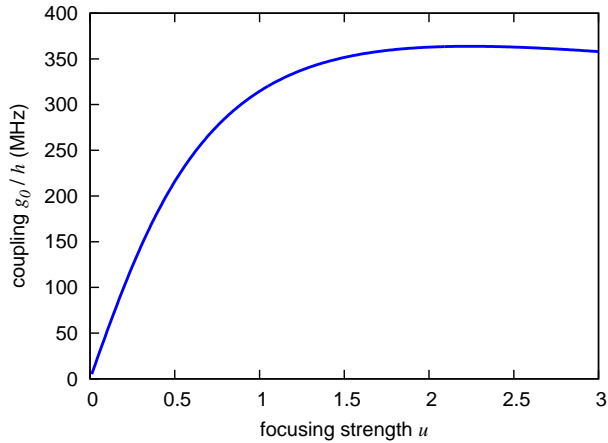}}
\caption{\label{fig:coupling}
  Coupling parameter $g_0$ as defined in the Jaynes-Cummings-Hamiltonian in
  eq.~(\ref{eq:jaynescummings}) for a two-level transition in $^{87}$Rb as a
  function of the focusing parameter $u$. Even for moderate focusing, large
  coupling constants exceeding the natural line width can be expected.
 }
\end{figure}

For the D2 transition in a $^{87}$Rb atom with a natural lifetime of
$\tau=26.25$\,ns and a cavity length of $L=10$\,mm, the expected coupling
constant $g_0$ is shown in figure~\ref{fig:coupling} as a function of the
focusing parameter $u$. One can see that even for moderate focusing
parameters, very large values for the coupling constant can be expected.

\subsection{Combining lenses and cavity}
A possible geometry with single meniscus elements for an atom-light interface
using both field enhancement by a cavity and strong focusing is shown
in figure~\ref{fig:cavityconcept}. The concave surfaces form the cavity, which
is almost concentric - and thus possibly difficult to align. 
We like to point out that for using our derivation of the focused light fields
\cite{Tey:09}, such cavity modes require that the mirror surfaces
be strictly spherical with a curvature radius $r_c$, and
$L\lesssim2r_c$. We approximated the focal position for the focused mode
as coinciding with the center of the wave front for the focusing calculations,
but this does not imply that such a cavity would be concentric.

Practical issues
aside, it has been known for a while that the transformation of the cavity
mode onto a collimated Gaussian beam can performed without any error
\cite{ibnsahl:984,Rashed:90}. For a focal length $f$, it can be shown that the shape of
the convex surface of this anaclastic lens is an ellipse with half axes
$fn/(n+1)$ in longitudinal, and $f\sqrt{(n-1)/(n+1)}$ in transverse direction
for a fixed refractive index $n$ of the lens.

Practically, it seems that a half-opening angle for the cavity mode of about
$\Theta_0\lesssim45^\circ$ can be achieved, limiting the focusing strength to about
$u_0=\tan\Theta_0\lesssim1$. Diffraction at the edges of this resonator will
limit the achievable finesse \cite{Fox:64}. Under the assumption that the
cavity eigenmode still
has a Gaussian angular profile, the round-trip diffraction loss $\epsilon$ at
the edges of the resonator for a mode with focusing strength $u$ for such a
cavity can be roughly estimated by 
\begin{equation}
\epsilon\approx e^{-2u_0^2/u^2}\,.
\end{equation}
For a focusing parameter $u\approx0.5$ and $u_0\approx1$, this loss is on the
order of $10^{-3}$. With this, a cavity finesse around 100 seems achievable,
and the strong coupling regime would be reachable with comparatively large
cavity geometries.

\section{Conclusion}
We have observed the back reflection of a coherent light field from a
single atom in near resonance with the light field, and could show that within
the limited localization of the atom in our experiment, we understand this in
a semiclassical picture for strongly focused light fields presented earlier
\cite{Tey:09}. We also presented a method of quantizing the fields in such a
strongly focused geometry. We find that, when using this geometry in a cavity
configuration with its discrete mode spectrum, a coupling constant of an atom
with this field much larger than the spontaneous decay time of an atom can be
obtained, even for cavities with a relatively large mirror separation. This
may imply a technical simplification of atom-photon interfaces to a point that
quantum information processing devices based on such interfaces
\cite{Cirac:97,Cirac:04} may become much more realistic.

\section*{Acknowledgements}
This work is supported by the National Research Foundation \& Ministry of
Education, Singapore.

\bibliography{atoms}


\end{document}